\documentclass[11pt,titlepage,a4paper]{article}
\usepackage{bozhomac}	
\usepackage{cite}

\title{\bfseries	\vspace*{-1.67890234in}
\vspace*{-07ex}
{
\begin{flushright}
	  \textbf{\large LANL xxx E-print archive No. gr-qc/9802058}\\[5ex]
\end{flushright}
}
{\huge On metric-connection~compatibility\\ and \\
	the signature change of space-time}
}

\vspace{2ex}

\author{
Bozhidar Z. Iliev
\thanks{Department Mathematical Modeling,
Institute for Nuclear Research and \mbox{Nuclear} Energy,
Bulgarian Academy of Sciences,
Boul. Tzarigradsko chauss\'ee~72, 1784 Sofia, Bulgaria}
\thanks{E-mail address: bozho@inrne.acad.bg}
}

\newlength{\bo}\newlength{\ho}\newlength{\up}\newlength{\down}       
\newlength{\middle}                                                  
\newcommand{\bozho}{\leavevmode\hbox{\slshape\bfseries               
\settowidth{\bo}{BO}\settowidth{\ho}{HO}\settowidth{\middle}{/}
\settoheight{\up}{BOZHO}\settodepth{\down}{/}
\addtolength{\up}{+0.15\up}\addtolength{\bo}{+\middle}
\rule[\up]{\bo}{0.15ex}\hspace{-\bo}BO
\hspace{+0.09em}\raisebox{+0.17\up}{/}
\hspace{-0.20em}\raisebox{+0.71\up}{$\bullet$}
\hspace{-0.33em}\hspace{-1.14\middle}\raisebox{-0.4\up}{$\bullet$}
\hspace{-0.30em}\addtolength{\down}{-0.41\down}
\addtolength{\ho}{+1.5\middle}\rule[-\down]{\ho}{0.15ex}
\addtolength{\ho}{-\middle}\hspace{-\ho}\hspace{+0.18em}
\raisebox{+0.17\up}{HO}}}                                            
\newcommand{\BOZHO}
{\bozho$^{^{\text{\textregistered}\,} \text{\texttrademark} }$}      
\date{
  \vspace{2.27ex}\ShortTitle{Metric-connection consistency
  			     and signature change}	\\[0.27ex]
  \vspace{3.27ex}
	\begin{tabular}{r@{$\colon\to~$}l}
  \vspace{0.09ex} Basic ideas	& April 13, 1997	\\[0.09ex]
  \vspace{0.09ex} Began		& April 15, 1997	\\[0.09ex]
  \vspace{0.09ex} Ended		& April 20, 1997	\\[0.09ex]
  \vspace{0.09ex} Revised	& September 1997, January 1998	\\[0.09ex]
 \vspace{0.09ex} Last update	& February 10, 1998	\\[0.09ex]
  \vspace{0.27ex} Produced	& \fbox{\today}	\\[0.27ex]
	\end{tabular} \\[1.27ex]
	\begin{tabular}{r@{$\colon~$}l}\large\mdseries\slshape
\vspace{0.27ex} LANL xxx archive server E-print No.& gr-qc/9802058\\[0.27ex]
	\end{tabular} \\[-0.27ex]
  \vspace{4.27ex}	{\Huge\BOZHO}	\\[4.27ex]
\vspace{0.27ex}\Subject{General relativity; Differential geometry}\\[2.27ex]
	\begin{tabular}{r@{\hspace{0.512em}}|@{\hspace{0.512em}}l}
\vspace{0.27ex} 	
&
\vspace{0.27ex}  
	\end{tabular} \\[1.27ex]
\vspace{0.27ex} 	\\[0.27ex]
}
\newcommand{\mat}[1]{\boldsymbol{#1}}	
\newcommand{\secref}[1]{Sect.~\ref{Sect#1}}

\begin{document}		

\renewcommand{\thefootnote}{\fnsymbol{footnote}} 
\maketitle			



\setcounter{equation} {0}

\begin{abstract}

	We discuss and investigate the problem of existence of
metric-compatible linear connections for a given space-time metric which is,
generally, assumed to be semi-pseudo-Riemannian. We prove that under
sufficiently general conditions such connections exist iff the rank and
signature of the metric are constant.  On this base we analyze possible
changes of the space-time signature.

\end{abstract}

\section{Introduction}
\label{Sect1}
\setcounter{equation} {0}

	It seems that for the first time in the works~\cite{sign5, sign2}
(see also~\cite{sign7,sign8}) were discussed space-time models with a
possible change of the signature of the space-time metric, called also
space-time signature or simply signature. At the beginning of the
nineties there appeared more often works on this
subject~\cite{sign16,sign17,sign9,sign14,sign18,sign1,sign4,sign3,sign23}.
Some of them study the generic mathematical structure of the space-time(s)
with changing
signature~\cite{sign17,sign10,sign12,sign6,sign18,sign30,sign26}, while
others deal with specific such
models~\cite{sign1,sign13,sign6,sign21,sign20,sign28,sign22}.
There are also articles investigating possible physical phenomena
that can happen if the signature
changes~\cite{sign3,sign4,sign1,sign15,sign19,sign22}.

%
	The main mathematical result of this work is a necessary and
sufficient condition for when a given semi-pseudo-Riemannian metric admits a
compatible (metric-compatible) with it linear connection. Freely speaking, we
can say that metric-compatible linear connections exist iff the rank and
signature of the metric are constant. (Notice, we consider
nondegenerate as well as degenerate metrics.) On this
rigorous base, analyzing the conditions responsible for the signature
constancy, we make conclusions for when the space-time signature can change.

	Above we mentioned in parentheses that the space-time metric can be
degenerate.  This requires some explanations since almost everywhere a
metric is defined as a non-degenerate Hermitian (resp. symmetric) quadratic
form in the complex (resp. real) case.
	For this purpose we shall fix first some concepts.
Following~\cite[pp.~66, 74]{Greub&et.al.-1}, we call $g$ a
\emph{pseudo-Riemannian} (resp.  \emph{pseudo-Hermitian})
\emph{metric} in a real (resp. complex) vector bundle
$\xi$ if it is symmetric (resp. Hermitian) bilinear (resp. linear in the
first and antilinear in the second argument) nondegenerate quadratic form on
the fibres of $\xi$. The pair $(\xi,g)$ is called pseudo-Riemannian (resp.
pseudo-Hermitian) vector bundle. If $g$ is positively defined, it is called
\emph{proper} or \emph{Riemannian};
otherwise it is called \emph{indefinite}. The same terminology
is transferred on differentiable manifolds~\cite[p.~273]{Bruhat} for which
$\xi$ is replaced with the corresponding bundles tangent to them, e.g. $(M,g)$
is a pseudo-Riemannian manifold if $g$ is a 2-covariant symmetric
nondegenerate tensor field on $M$. If in the above definitions is omitted the
nondegeneracy condition, to the corresponding concepts is added the prefix
`\emph{semi-}'~\cite{Rosenfel'd} (see also the references
in~\cite[remark~8 to chapter~VI (p.~508)]{Rosenfel'd}); for instance,
a \emph{semi-pseudo-Riemannian metric} on a manifold $M$ is a 2-covariant
symmetric tensor field on $M$ (see also~\cite[e.g., the articlle
``semi-pseudo-Riemannian space'']{Mathenedia-4}).

	\begin{rem}{1.1}\small
	A different, but analogous, terminology is used in the theory of
linear partial differential equations of second
order~\cite{Tricomi,MoiseevEI,FriedrichsKO1958,FriedrichsKO1970,
SmirnovMM,Oleinik&Rodkevich,BerezanskiiYuM}.
The type of the equation
\(
\sum_{i,j=1}^{n}A_{ij}(x) \partial^2 u/\partial x^i\partial x^j
+
F(x, u, \partial u/\partial x^1, \ldots, \partial u/\partial x^n)
=0
\),
 $x=(x_1,\ldots,x_n)\in\mathbb{R}^n$ is determined by the eigenvalues of the
matrix $A(x):=[A_{ij}(x)]$~\cite[chapter~8, \S~2]{MikhlinSG} which plays the
role of a metric. This equation is of type $(p,q,r)$ at
$x\in\mathbb{R}^n$ if at $x$ the matrix $A(x)$ has $p$ positive, $q$
negative, and $r$ zero eigenvalues.%
\footnote{%
This classification can be extended also on non-linear partial differential
equations of second order~\cite{MikhlinSG}.%
}
The type is called elliptic, parabolic, or hyperbolic at a given point if it
has respectively the form
$(m,0,0)$ or $(0,m,0)$,
$(m-1,0,1)$ or $(0,m-1,1)$,
$(m-1,1,0)$ or $(1,m-1,0)$
at that point. Using this terminology, the (proper) Riemannian and Lorentzian
metrics %
\footnote{%
An indefinite pseudo-Riemannian metric with exactly one positive or negative
eigenvalue is called Lorentzian.%
}
can be called respectively (globally) elliptic and (globally) hyperbolic
metrics; there are no `parabolic metrics' between the pseudo-Riemannian ones.
The most widely investigated case is when equation's type is constant in the
region where it is considered, e.g. in the whole space; in our analogy the
pseudo-Riemannian metrics with constant signature correspond to (part of)
such equations. If the equation's type changes from point to point, it is
said to be of \emph{mixed
type}~\cite{Kuzmin'AG,GuChaohao,SmirnovMM,Tricomi,MoiseevEI}.%
\footnote{%
A classical example of such an equation is the Tricomi equation
$ \partial^2u/\partial^2y + y \partial^2u/\partial^2x = 0 $.%
}
The `metrical' analog of these equations are pseudo-Riemannian metrics with
changing (from point to point) signature, therefore they can be called
\emph{mixed (pseudo-)Riemannian} metrics. For such metrics, because of their
non-degeneracy, always exist points at which they are not continuous (or
smooth). However, the equations of type $(p,q,r)$ with $r\ge1$ do not have
analogs between pseudo-Riemannian metrics as the latter are, by definition,
nondegenerate. In this context, it is not difficult to see that there is a
full one-to-one correspondence between the classification by type of the
(linear) partial differential equations of second order and the class of
semi-pseudo-Riemannian metrics. For instance, now there are parabolic
metrics, defined as semi-pseudo-Riemannian ones with defect 1, i.e. with
exactly 1 vanishing eigenvalue; also one can freely investigate continuous
(or smooth) \emph{mixed semi-pseudo-Riemannian} metrics, i.e. ones with
changing signature and points at which they are degenerate, as they are always
somewhere degenerate, etc.
	Ending this long remark, we want to emphasize on the fact that
the revealed analogy between semi-pseudo-Riemannian metrics and the
(classification of the) (linear) partial differential equation of second
order is another argument for the investigation of degenerate metrics.
	\end{rem}

	For the general considerations of problems concerning signature
changes and metric-connection compatibility one should work with
semi-pseudo-Riemannian (resp. semi-pseudo-Hermitian) metrics instead of
conventional pseudo-Riemannian (resp. pseudo-Hermitian) ones in the real
(resp.  complex) case.
	This allows to be retained the concept of a global
metric in some natural cases in which the conventional definition breaks
down. For instance, let a manifold $M$ be divided into several (not less then
two) regions $U_i,\ i=1,2,\ldots$
Let on each $U_i$ be given a
pseudo-Riemannian smooth metric $g_i$ and there to exists
a symmetric tensor field $g$ on $M$ whose restriction on $U_i$
is exactly $g_i$, $g|U_i=g_i$. It is natural to call  $g$ a (global)
`pseudo-Riemannian metric' on $M$. But is this `definition' correct in the
conventional sense? Yes, if the signatures of all $g_i$ are equal and $g$ is
smooth.  But if at least two (local) metrics, say $g_i$ and $g_j$, have
different signatures, then for $g$ must exist regions (possibly with
dimension less then $\dim M$) on which it is degenerate or/and discontinuous.
In particular, if $g$ is continuously defined on $M$, then it must be
degenerate somewhere and, consequently, $g$ is not a pseudo-Riemannian
metric. Looking over the literature, we see that the last situation is not
an exceptional one, on the contrary, it is the case usually
considered~\cite{sign1,sign4,sign15,sign16,sign21}.  Moreover, often $g$ is
called (intuitively) a metric in such cases nevertheless that it is a
degenerate tensor field!  The above considerations force us to use
semi-pseudo-Riemannian metrics instead of pseudo-Riemannian ones. So, in this
paper we shall omit the nondegeneracy condition in the metric's definition.
Hence, the metrics investigated here can be non-degenerate as well as
degenerate ones. In this way we achieve a uniform description of problems
which otherwise have to be investigated separately.

	In \secref{2} we present some preliminary mathematical material on
which our investigation rests. In \secref{3} the problem for when a given
metric admits a metric-compatible linear connection is rigorously analyzed.
Here results concerning the more general problem on metric-compatible linear
transports along paths are given too. In \secref{4} we make conclusions on
the possible changes of the space-time signature.  \secref{conclusion} closes
the paper with some concluding remarks.

\section {Linear transports along paths and \\
		their compatibility with a fibre metric }
\label{Sect2}
\setcounter{equation} {0}

	In this section we recall some facts concerning linear transports
along paths in vector bundles~\cite{f-LTP-general} and briefly review the
problem of their compatibility (consistency) with a metric on the
bundle~\cite{f-LTP-metrics}.

	Let $(E,\pi,M)$ be a complex vector fibre bundle with base $M$, total
space $E$, and projection $\pi :E\to M$. The fibres $E_x:=\pi^{-1}(x)\subset
E$, $x\in M$, are isomorphic vector spaces, i.e. there exists a vector space
${\mathcal E}$ and linear isomorphisms $l_x$, $x\in M$  such that
$l_x:E_x\to {\mathcal E}$. We do not make any assumptions on the
dimensionality of $(E,\pi,M)$, i.e. ${\mathcal E}$  can be finite as well
as infinite dimensional.


	By $J$ and $\gamma:J\to M$ we denote a real interval and a path in
$M$, respectively.

	A \emph{$\mathbb{C}$-linear transport (L-transport) along paths}
in $(E,\pi,M)$ is a map
$L:\gamma \mapsto L^{\gamma }$, where
$L^{\gamma }:(s,t)\mapsto L^{\gamma }_{s\to t}, \  s,t\in J$
is the (L-)transport along $\gamma$, and
$L^{\gamma }_{s\to t}:\pi ^{-1}(\gamma (s)) \to \pi ^{-1}(\gamma (t))$,
called (L-)transport along $\gamma$ from $s$ to $t$,
satisfies the equalities
\begin{eqnarray}
&
 L^{\gamma }_{t\to r}\circ L^{\gamma }_{s\to t}=L^{\gamma }_{s\to r},
\quad r,s,t\in J,
& \label{2.1} \\
&   L^{\gamma }_{s\to s}= {id}_{\pi ^{-1}(\gamma (s))}, \quad s\in J,
& \label{2.2} \\
&   L^{\gamma }_{s\to t}(\lambda u +\mu v)=
\lambda L^{\gamma }_{s\to t}u +
\mu L^{\gamma }_{s\to t}v, \quad \mu,\lambda\in \mathbb{C},
\quad u,v\in\pi ^{-1}(\gamma (s)).
& \ \ \ \label{2.3}
\end{eqnarray}
Here $id_N$ denotes the identity map of a set $N$.
The general form of $L^{\gamma }_{s\to t}$ is
\begin{equation}
 L^{\gamma }_{s\to t}={\left( F^{\gamma }_{t} \right)}^{-1}\circ
 F^{\gamma }_{s}, \quad s,t\in J       \label{2.31}
\end{equation}
with $F^{\gamma }_{s}:\pi ^{-1}(\gamma (s)) \to Q, \  s\in J$, being
one-to-one linear maps onto one and the same (complex) vector space Q.

	Equations~(\ref{2.1}) and~(\ref{2.2}) imply
	\begin{equation}	\label{-1}
\left( L^{\gamma }_{s\to t}\right)^{-1}=L^{\gamma }_{t\to s}.
	\end{equation}

	According to~\cite[theorem~3.1]{f-TP-parallelT} the
set of (linear) transports which are diffeomorphisms and satisfy the locality
and reparametrization conditions%
\footnote{%
I.e.\  respectively $ L^{\gamma }_{s\to t}\in
\mathrm{Diff}(\pi^{-1}(\gamma (s)),\pi^{-1}(\gamma (t)))$,
$ L^{\gamma |J^\prime }_{s\to t}=L^{\gamma}_{s\to t}$ for $s,t\in J^\prime $,
with $J^\prime $ being a subinterval of $J$, and
$ L^{\gamma\circ\tau}_{s\to t}=L^{\gamma}_{\tau(s)\to \tau(t)} $,
$s,t\in J''$ with $\tau$ being a 1:1 map of an $\mathbb{R}$-interval  $J''$
onto $J$.%
}
are in one-to-one correspondence with the (axiomatically defined)
(linear) parallel transports (along curves). So, the conventional parallel
transport along $\gamma$ from $\gamma (s)$ to $\gamma (t)$, assigned to a
linear connection, is a standard realization of the general (linear)
transport $L^{\gamma }_{s\to t}$.

	Let $g$ be a semi-pseudo-Hermitian fibre metric on $(E,\pi,M)$.
This means
that $g:x\mapsto g_x$ with $g_x:E_x\times E_x\to {\mathbb C}$,
$x\in M$, being Hermitian forms~\cite{Rosenfel'd,K&N-1},
i.e.  $g_x$ are ${\mathbb C}$-linear
in the first argument and ${\mathbb C}$-antilinear in the second argument
Hermitian maps.
\begin{defn}[{[cf.~\protect{\cite[definition~4.1]{f-LTP-metrics}}]}]{3.1}
A fibre semi-pseudo-Hermitian metric $g$ and a linear transport $L$ are
called compatible (resp. along $\gamma$) if $L$ preserves the scalar product
defined by $g$, i.e.

\begin{equation}
g_{\gamma (s)}=g_{\gamma (t)}\circ\left( L^{\gamma }_{s\to t}\times
L^{\gamma }_{s\to t}\right),\quad s,t\in J \label{2.7}
\end{equation}
for all (resp. the given) $\gamma$.
\end{defn}

	In~\cite{f-LTP-metrics} different results concerning the
compatibility of linear transports along paths and fibre metrics can be
found. They are prove in~\cite{f-LTP-metrics} only in the finite dimensional
case, i.e. for $\dim\mathcal{E}<\infty$. However, some of them remain valid
also in the infinite dimensional case. For instance, proposition~4.3
of~\cite{f-LTP-metrics} is easily shown to be insensitive to the bundle's
dimensionality, i.e. for any ($\mathbb{C}$-)linear transport along paths
there exist consistent with it Hermitian fibre metrics along any fixed path.
The general form of these metrics is given
by~\cite[equation~(4.8)]{f-LTP-metrics}. The global version, viz. along
arbitrary paths, of this statement is not always true. In the finite
dimensional case it is expressed by~\cite[proposition~4.6]{f-LTP-metrics},
which \emph{mutatis mutandis} holds and for an infinite dimension.

	Below we are going to use the following result
(cf.~\cite[proposition~4.4]{f-LTP-metrics}).

	\begin{prop}{2.1}
Let $(E,\pi,B)$ be a finite  dimensional complex vector bundle endowed with a
pseudo-Hermitian fibre metric $g$. A necessary and sufficient
condition for the existence of a compatible (resp. along a fixed path) with
$g$ ($\mathbb{C}$-)linear transport along paths (resp. along the given path)
is the independence of the signature of $g$ of the point of
$B$ (resp. of the fixed path) at which it is calculated.
	\end{prop}

	We want to make several comments on this result which is a
simple reformulation of~\cite[proposition~4.4]{f-LTP-metrics}.

	Proposition~\ref{prop2.1} holds also for pseudo-Riemannian metrics
(see~\cite[proposition~2.4]{f-LTP-metrics}) as they are evident special case
of the pseudo-Hermitian ones~\cite{K&N-2}.

	The validity of proposition~\ref{prop2.1} is limited to finite
dimensional vector bundles and cannot be generalized to infinite dimensional
ones. There are two main reasons for this. On one hand, in its formulation is
involved the notion of signature of Hermitian
forms~\cite[section~2.12]{Lankaster-matrices}, which is generically a finite
dimensional concept since it is (usually - see below) defined as the
difference between the number of positive and number of negative eigenvalues
of a form, i.e. between its positive and negative (inertia)
indexes~\cite[p.~334]{Gantmacher-matrices-I}.
On the other hand, the proof of the discussed proposition
(see~\cite{f-LTP-metrics}) uses \emph{essentially} the ((Jacobi-)Sylvester)
law of inertia for Hermitian (or symmetric, in the real case) quadratic forms
(see, e.g.,~\cite[section~2.12]{Lankaster-matrices}
and~\cite[p.~297 and p.~334]{Gantmacher-matrices-I}).
The law of inertia can be generalized for (degenerate of not) Hermitian
(resp. symmetric) forms over ordered fields. However this is possible only
for finite dimensional vector spaces~\cite[chapter~12,
\S~90]{Waerden-algebra-II}.

	Since the law of inertia has a form valid also for degenerate
Hermitian forms~\cite[section~2.12]{Lankaster-matrices},
the proofs of propositions~2.4 and~4.4 of~\cite{f-LTP-metrics}
can be mended \textit{mutatis mutandis} by its help in such a way that they
remain true for degenerate metrics too. This leads us to the following
generalization of proposition~\ref{prop2.1}.

	\begin{prop}{2.2}
Let $(E,\pi,B)$ be a finite  dimensional complex vector bundle endowed with a
semi-pseudo-Hermitian fibre metric $g$. There exists a
compatible (resp. along a fixed path) with $g$ ($\mathbb{C}$-)linear
transport along paths (resp. along the given path) if and only if the
rank and signature of $g$ are independent of the point of $B$ (resp. of
the fixed path) at which they are calculated.
	\end{prop}

	Ending this section, we pay attention to the definition of metric's
signature. At $x\in B$ a metric  $g$  is represented by Hermitian form
$g_x\colon E_x\times E_x\to\mathbb C$. Let $p(x)$ and $q(x)$ be its,
respectively, positive and negative (inertia) indexes, which are equal to (or
can be defined as) the number of positive and negative, respectively,
eigenvalues of $g_x$, i.e. of the matrix representing $g_x$ in some local
bases. Mathematically the signature of $g$ at $x$  is defined as the pair
$(p(x),q(x))$ or, more often, as the number
$s(x)=p(x)-p(x)$~\cite{Lankaster-matrices}. The last definition will be used
in this work.%
\footnote{%
	The law of inertia states that the numbers $p(x),\ q(x),\ s(x)$, and
the rank $r(x)$ of $g_x$  are invariants that are independent of the
local bases by means of which they are
determined~\cite{Lankaster-matrices,Gantmacher-matrices-II}.%
}
	In the physical literature 
signature of a nondegenerate metric is called the put in parentheses
sequence $(++\cdots --)$ of $p(x)$ plus signs and $q(x)$ minus signs,
corresponding to the eigenvalues of the metric, in the order in which they
appear in the (standard) diagonal form of the matrix representing $g_x$ in a
some local basis. This definition of the signature, that can be called
physical, will not be used here.

\section[Which metrics admit metric-compatible connections?]
	{Which metrics admit \\ metric-compatible connections?}
\label{Sect3}
\setcounter{equation} {0}

	Most of the modern nonquantum gravity
theories~\cite{Gronwald/Hehl-96,MTW} are constructed by means of two basic
geometrical structures over the space-time (real) manifold $M$, viz. a
pseudo-Riemannian metric $g$ and a linear connection (covariant derivative)
$\nabla$  which is a specific derivation of the tensor algebra over
$M$~\cite{K&N-1}. So, the metric is supposed to be a nondegenerate,
symmetric, and two times contravariant tensor field~\cite{K&N-1,K&N-2}, i.e.
a nondegenerate section of the symmetric tensor bundle of type $(0,2)$ over
$M$. These structures are called \emph{compatible} (or consistent) on $M$ if
$g$ is of class $C^1$ and
	\begin{equation}	\label{3.1}
\nabla_X (g) = 0           
	\end{equation}
for any vector (field) $X$. In this case the connection $\nabla$  is called
\emph{metric-compatible} (with the given metric on $M$). Examples of
gravitational theories based on a metric-compatible connections are general
relativity, Einstein-Cartan ($U_4$) theory, and Einstein teleparallelism
theory~\cite{Gronwald/Hehl-96}.

	Given a ($C^1$) pseudo-Riemannian metric $g$, it is
known~\cite[chapter~IV, \S~2]{K&N-1} that there always exists a
metric-compatible with it linear connection (which is unique in the
torsionless case). Here a natural question arises: what properties of $g$ are
responsible for this existence? Looking over the proof(s) of the existence
of metric-compatible connection(s), one finds, at first sight, that they
essentially use that the metric is nondegenerate and of class of smoothness
not less than $C^1$. So, if one of these conditions breaks, one can expect a
nonexistence of metric-compatible connection(s). However, as we shall see
below, these are not the primary causes for such nonexistence. To examine
this problem in details, we shall reformulate~(\ref{3.1}) in terms of
parallel transports (translations).

	Let $\gamma\colon[0,1]\to M$ be a $C^1$ path in $M$ and
$g\colon x\mapsto g_x,\ x\in M$, with
$g_x\colon T_x(M)\times T_x(M)\to\mathbb R$. Here $T_x(M)$ is the space
tangent to $M$ at $x$. The linear connection $\nabla$ can equivalently be
described by the concept of parallel transport $\tau$ along
paths (see~\cite[section~5.2]{Bishop/Crittenden} and~\cite{K&N-1}).%
\footnote{\label{fn-nabla}%
If $\nabla$ is given, one can define
\(
\tau\colon\gamma\mapsto\tau^\gamma\colon
T_{\gamma(0)}(M)\to T_{\gamma(1)}(M)
\)
by \(\tau^\gamma(A_0)=A(1)\), where \(A_0\in T_{\gamma(0)}(M)\)
and the $C^1$ vector field $A$ is given on $\gamma([0,1])$
via the initial-value problem
\(\left.(\nabla_V A)\right|_{\gamma([0,1])} = 0\), \(A(0)=A_0\).
Here $V$ is a vector field which on $\gamma([0,1])$ reduces to the vector
field tangent to $\gamma$.%
}

	Let $\tau\colon\gamma\mapsto\tau^\gamma$ with $\tau^\gamma$ being the
parallel transport along $\gamma$ defined by $\nabla$
(see footnote~\ref{fn-nabla} on page~\pageref{fn-nabla}).
The compatibility condition~(\ref{3.1}) is equivalent to the requirement that
$\tau$ preserves the defined via $g$ inner products along any path
$\gamma$~\cite{Bishop/Crittenden,K&N-1}, i.e.
	\begin{equation}	\label{3.2}
g_{\gamma(0)} = g_{\gamma(1)} \circ (\tau^\gamma\times\tau^\gamma).
	\end{equation}
Since any parallel transport along paths (curves), like $\tau$, is a (linear
in our case) transport along paths~\cite{f-TP-parallelT}, we see that
equation~(\ref{3.2}), and hence~(\ref{3.1}), is equivalent to~(\ref{2.7}) for
$s=0,\ t=1 \text{ and } L=\tau$. Consequently, when analyzing the conditions
for the validity of~(\ref{3.1}), we can apply proposition~\ref{prop2.2}.

	So, let $g$ be a semi-pseudo-Riemannian metric on $M$, i.e. $g\colon
x\mapsto g_x$ where $g_x\colon T_x(M)\times T_x(M)\to\mathbb R,\ x\in M$ are
symmetric bilinear real quadratic forms. Under what additional conditions on
$g$ there is a parallel transport $\tau$ such that the compatibility
equation~(\ref{3.2}) holds?

	Let $r(x)\text{ and } s(x)$ be respectively the rank and signature of
$g(x)$ at $x$. Proposition~\ref{prop2.2}, when applied to the bundle tangent
to $M$, shows that a necessary and sufficient condition for the validity
of~(\ref{3.2}) for some transport $\tau$ along paths is the independence of
$r(x)$ and $s(x)$ of the point $x$ at which  they are calculated, i.e.
$r(x)=\const \text{ and } s(x)=\const$ for every $x\in M$.

	Therefore, if $g$ has a constant rank and signature, there exists a
linear transport $\tau$ along paths consistent with it. However, is this
transport a parallel one, i.e does there exists a linear connection $\nabla$
for which $\tau$ is a parallel transport (see footnote~\ref{fn-nabla} on
page~\pageref{fn-nabla})?

	Let $\mat{F}_{s}^{\gamma}$ be the matrix corresponding to the map
${F}_{s}^{\gamma}$ (see~(\ref{2.31})) in some local bases.
From~\cite[equation~(4.9)]{f-LTP-general} we know that the matrix of the
coefficients of a linear transport $L$ along paths, if it has a $C^1$
dependence on its parameters, is
	\begin{equation}	\label{3.3}
\mat{\Gamma}(s;\gamma) := \left[{\Gamma}_{\ j}^{i}(s;\gamma)\right] =
\left(\mat{F}_{s}^{\gamma}\right)^{-1}
\frac{d\mat{F}_{s}^{\gamma}}{ds}.
	\end{equation}
The considerations in~\cite[section~5]{f-LTP-general} show that $L$ is the
parallel transport corresponding to a linear connection with local
coefficients ${\Gamma}_{\ jk}^{i}(x)$ iff
	\begin{equation}	\label{3.4}
\mat{\Gamma}(s;\gamma) =
\sum_{k=1}^{n} \mat{\Gamma}_k(\gamma(s))\dot\gamma^k(s)
	\end{equation}
where
$\mat{\Gamma}_k(x) := \left[\Gamma_{\ jk}^{i}(x)\right]_{i,j=1}^{n}$,
$n=\dim M$ and $\dot\gamma(s)$ is the vector tangent to $\gamma$ at
$\gamma(s)$.

	Consequently, a necessary condition for  $\tau^\gamma$  to be a
parallel transport, assigned to some $\nabla$, along $\gamma$ is $\gamma$ to
be a $C^1$  path.

	So, let $\gamma$  be a $C^1$ path. Moreover, if $r(x)=n:=\dim M$, i.e.
if $g$  is nondegenerate, we can reconstruct $\nabla$  from $g$  in the well
known way~\cite{K&N-1,MTW}. However, is it possible to be found a
metric-compatible connection $\nabla$ for a degenerate metric $g$, i.e. for
$r(x)<n$?  Surprisingly the answer to this question is positive. To prove
this we need the following generalization
of~\cite[proposition~2.5]{f-LTP-metrics}.

	\begin{prop}{3.1}	
Let in the $n$-dimensional, $n<\infty$, real vector bundle $(E,\pi,M)$ be
given a fibre semi-pseudo-Riemannian metric $g$ with constant rank and
signature along every (resp.  some fixed) path $\gamma\colon J\to M$. Let
\(
\{ \{e_i(\gamma(s)) : i=1,\dots,n\}\text{ - basis in $E_{\gamma(s)}$} \}
\)
be a field of bases along $\gamma$ in which $g$ is represented by the matrix
$\mat{G}(\gamma(s)) = \left[ g_x(e_i(x),e_j(x))\right]_{x=\gamma(s)}$.
Suppose $\mat{D}(\gamma(s))$ is a nondegenerate (real) matrix transforming
$\mat{G}(\gamma(s))$ to a diagonal form by means of congruent
transformation:
	\begin{equation}	\label{3.5}
\mat{D}^\top(\gamma(s))\mat{G}(\gamma(s))\mat{D}(\gamma(s)) =
\mat{G}_0(\gamma(s)) := \diag(d_1(\gamma(s)),\dots,d_n(\gamma(s)))
	\end{equation}
where $\top$  means matrix transposition,  $p$ (=the number of positive
eigenvalues of $g$) of the real numbers
$d_1(\gamma(s)),\ldots,d_n(\gamma(s))$ are positive,
$q=r-p$(=the number of negative eigenvalues of $g$) of them are negative, and
the remaining \mbox{$n-r$}(=the number of zero eigenvalues of $g$) of them
are equal to zero. Then the set of \textsl{all} linear transports along paths
compatible with $g$ along every (resp. the given) path $\gamma$ is
described via the decomposition~\eref{2.31} in which the matrix of the map
$F_{s}^{\gamma}$ has the form
	\begin{equation}	\label{3.6}
\mat{F}_{s}^{\gamma} =
\mat{B}(\gamma)\mat{Z}(s;\gamma)(\mat{D}(\gamma(s))^{-1}
	\end{equation}
for every (resp. the given) path $\gamma$. Here $\mat{B}(\gamma)$ is a
nondegenerate $n\times n$ matrix function of $\gamma$ and $\mat{Z}(s;\gamma)$
is any nondegenerate $n\times n$ matrix function of  $s$ and $\gamma$
satisfying the equality
	\begin{equation}	\label{3.7}
\mat{Z}^\top(s;\gamma)\mat{G}_0(s)\mat{Z}(s;\gamma) = \mat{G}_0(s).
	\end{equation}
        \end{prop}	

	\begin{rem}{1}
By renumbering of the vectors of the local bases and renormalizing $\mat{D}$
and $\mat{Z}$ we can choose $\mat{G}_0$ in the form
	\begin{equation*}	
\mat{G}_0(\gamma(s)) = \mat{G}_{p,q,n-r} :=
\diag(
\underbrace{+1,\dots,+1}_{\text{$p$ times} },
\underbrace{-1,\dots,-1}_{\text{$q$ times} },
\underbrace{0,\dots,0}  _{\text{$(n-r)$ times} }
).
	\end{equation*}
In this case $\mat{Z}$ could be called a semi-pseudo-orthogonal matrix of
type $(p,q,n-r)$ or of type $(p,q)$ and defect $n-r$ (cf. the corresponding
terminology concerning metrics~\cite{Rosenfel'd}).
	\end{rem}

	\begin{rem}{2}
This proposition has an evident generalization for complex fibre bundles. In
this case $g$ is a semi-pseudo-Hermitian metric and the matrix transposition
has to be replaced with Hermitian conjugation.
	\end{rem}

	\begin{proof}[Proof.]
\emph{Mutatis mutandis} this proof is an exact copy  of the one
of~\cite[proposition~2.5]{f-LTP-metrics}. We have simply to make use of the
new definition of $\mat{D}$ (via~\eref{3.5}) whose existence is proved, for
instance, in~\cite[section~2.12]{Lankaster-matrices}.
	\end{proof}

	Combining~\eref{3.3} with proposition~\ref{prop3.1}, we conclude that
a linear $C^1$ transport compatible with a metric $g$ has
coefficients whose matrix is of the form
	\begin{equation}	\label{3.8}
	\begin{split}
\mat{\Gamma}(s;\gamma)	&=
\left\{ \mat{Z}(s;\gamma) \mat{D}^{-1}(\gamma(s)) \right\}^{-1}
\frac{d}{ds} \left\{ \mat{Z}(s;\gamma) \mat{D}^{-1}(\gamma(s)) \right\}   \\
			&=
\mat{D}(\gamma(s)) \mat{Z}^{-1}(s;\gamma)
\frac{d \mat{Z}(s;\gamma)}{ds} \mat{D}^{-1}(\gamma(s)) -
\frac{d \mat{D}(\gamma(s))}{ds} \mat{D}^{-1}(\gamma(s)).
	\end{split}
	\end{equation}
Here have we supposed $\mat{Z}(s;\gamma)$ and $\mat{D}(x)$ to be of class
$C^1$ with respect to $s$  and $x$ respectively, i.e. we have assumed that
the metric $g$ is of class $C^1$.

	Now comparing~\eref{3.4} and~\eref{3.8}, we see that a linear
transport with coefficients given by~\eref{3.8} is a parallel transport
for some linear connection $\nabla$ iff
	\begin{equation}	\label{3.9}
\mat{Z}(s;\gamma) = \widetilde{\mat{Z}}(\gamma(s)),
	\end{equation}
i.e. iff the matrix $\mat{Z}(s;\gamma)$  depends only on the point
$\gamma(s)$ but not on $s$ and $\gamma$ separately. Besides, in this case
$\nabla$ has local coefficients given by
	\begin{equation}	\label{3.10}
\mat{\Gamma}_k(x) = [ {\Gamma}_{\ jk}^{i}(x) ] =
\mat{D}(x) \widetilde{\mat{Z}}^{-1}(x)
\frac{\partial \widetilde{\mat{Z}}(x)}{\partial x^k} \mat{D}^{-1}(x) -
\frac{\partial \mat{D}(x)}{\partial x^k} \mat{D}^{-1}(x)
	\end{equation}
where $\{x^k\}$ are local coordinates in a neighborhood of $x\in M$.

	Notice, by~\eref{3.10} a necessary condition for $L$ to be a
parallel transport for a metric-compatible connection is the metric $g$  to
be of class $C^1$.

	The overall above discussion can be summarized in the
following theorem in which the above results are slightly generalize
by introducing a set $U\subseteq M$.

	\begin{thm}{3.1}
Let on $U\subseteq M$ the semi-pseudo-Riemannian metric $g\colon x\mapsto
g_x$ be defined by the symmetric $C^1$ quadratic forms
$g_x\colon T_x(M)\times T_x(M)\to\mathbb R,\ x\in U$. A necessary and
sufficient condition for the existence of a metric-compatible with $g$ linear
connection in $U$ is the independence of the rank $r(x)$ and signature
$s(x)$ of $g_x$ of the point $x\in U$ at which they are calculated. Given $g$
with these properties, the set of all such connections is selected
via~\eref{3.10}. Moreover, the set of parallel transports corresponding to
these connections coincides with the set of linear smooth ($C^1$) transports
along smooth ($C^1$) paths which transports are compatible with $g$.
	\end{thm}

\section{When the space-time signature can change?}
\label{Sect4}
\setcounter{equation} {0}

	It is well known that general relativity is based on a
pseudo-Riemannian metric of class $C^2$  over a 4-manifold $V_4$~\cite{MTW}
and the compatible with it torsionless linear connection, called Riemannian
or Levi Civita's connection~\cite{Bishop/Crittenden,MTW}, whose coefficients
are the Christoffel symbols formed from the metric~\cite{MTW,K&N-1}.
By theorem~\ref{thm3.1} this metric must have a constant signature,
conventionally assumed to be +2 or -2,  or, in physical terms,
$(-+++)$ or $(+---)$ respectively~\cite{MTW}.
Thus, if one wants to build axiomatically general
relativity, it is sufficient to suppose the existence of a torsionless
metric-compatible connection or a constant signature of the
pseudo-Riemannian metric.%
\footnote{%
In any one of these possibilities the concrete signature can be fixed via the
equivalence principle~\cite{MTW,f-PE-P?}.%
}
Therefore in general relativity the signature is constant over the whole
space-time.

	In all known to the author (nonquantum) gravitational theories, the
metrics, if any, are pseudo-Riemannian and of class $C^1$ or $C^2$
~\cite{MTW,Gronwald/Hehl-96}. According to theorem~\ref{thm3.1}, all such
theories which use a metric-compatible connection, e.g. Einstein-Cartan
and Einstein teleparallelism theories, must have a constant metric's
signature. In other theories, such as Weyl's and metric-affine gravity,
which are based on metric-incompatible connections, the signature is
allowed, at least in principle, to change from point to point. Nevertheless
that the last possibility potentially exists, it is not realized until now
into a consistent gravitational theory that can stand experimental
checking~\cite{Gronwald/Hehl-96,MTW,Hehl-95-review}.

	In the literature can be found papers devoted to the
mathematical structure and possible physical events in space-time(s) with
changing
signature~\cite{sign1,sign3,sign8,sign9,sign12,sign17,sign18,sign20,sign21}.
Most of them are based on modifications of general
relativity~\cite{sign1,sign8,sign13,sign14,sign15,sign16,sign17,sign18}.
In such models the metric is globally only a symmetric quadratic form for
which can exist sets on which it is degenerate or/and not differentiable or
even discontinues. Excluding these peculiar sets, on the remaining parts
(sets) of the space-time the metric is assumed, as usual, to be a symmetric,
nondegenerate, and smooth ($C^1$ or $C^2$) quadratic form. On these latter
sets, which can be called regular for the metric, is supposed to be valid
general relativity. Consequently, by theorem~\ref{thm3.1}, the signature
of the metric is constant on them, but on the different sets it can be a
different constant. On the sets on which the metric is degenerate the
signature also can change from one to another set, depending on the number of
zero eigenvalues of the metric on them. (Notice, the signature may have
different values on the last sets and on the regular ones.) If on these sets,
if any, the metric is still assumed to be symmetric, with a constant rank on
them, and smooth ($C^1$ or $C^2$), then, due to theorem~\ref{thm3.1}, on them
can be suggested to be valid general relativity. Since in the last case the
three index Christoffel symbols (of second kind) do not exist, the
connection coefficients have to be calculated by~\eref{3.10}. At the end,
there can be sets on which the smoothness of the metric breaks.%
\footnote{%
On these sets, usually, the metric is also degenerate as they often play a
role of boundaries (sets of zero measure) between sets of the previous
class~\cite{sign1,sign13,sign15,sign18,sign23,sign3}.%
}
On them theorem~\ref{thm3.1} is not valid, metric-compatible connections do
not exist, and, as a whole, on them a version of general relativity cannot be
constructed. Hence these sets, if any, are the most probable ones on which
the space-time signature may change.

\section {Conclusion}
\label{Sectconclusion}
\setcounter {equation} {0}

	In this work we have investigated the problem for possible change of
the space-time signature from the view-point of existence of
metric-compatible connections. Its main moral is:
\emph{%
on some space-time region there exists a metric-compatible connection if and
only if the corresponding (degenerate or not) $C^1$
semi-pseudo-Riemannian metric has a constant signature and rank in this
region.%
}
For a globally defined metric with changing signature and, possibly, rank
there is not a globally consistent with it linear connection, but on some
subsets of the space-time such connection may exist.

	When speaking about the space-time in this work, we implicitly
suppose to be dealing with a classical, not quantum, theories of gravity in
which it is a four-manifold. In quantum theories, like supergravity and
string ones, the multidimensional, greater then four,
character of space-time is accepted. In such theories there are possible
(and necessary) transitions between geometries with changing space-time
dimensions when the metric's signature and rank are
variable~\cite{sign23,sign24,sign25,sign27}.
Since theories of this kind are based on mathematical structures different
from linear connections, they do not fall in the subject of the present
paper.

	In connection with the metric-affine gravitational
theories~\cite{Gronwald/Hehl-96,Hehl-95-review}
the following question may arise.
Given on the space-time (or on some its subset) an affine connection, is
there a metric with which it is metric-compatible? This problem is, in some
sense, opposite to the one investigated in the present work and will be
considered elsewhere. Here we want only to mention that it can be solved
completely by using the methods of this paper
and~\cite[propositions~2.3 and~2.6]{f-LTP-metrics} which can easily be
generalized to describe degenerate metrics too.

	Practically all of the mathematical results of this work,
concerning the real case, can be generalized to the complex one. So, if
required, they can be reformulated in terms of (complex or real) manifolds
and (semi-)Hermitian (degenerate or not) metrics on them.

	And a last remark. At a classical level, we know that there are three
space and only one time dimension. This fact is reflected in the accepted,
e.g. in general relativity, space-time signature. We also know from
experience that under normal conditions, such as on the Earth or in the Solar
system, the space-time signature is constant. This observation, combined with
the results of the above investigation, leads to the conclusion that at
present there are not experimental results which have to be described via
space-time(s) with changing signature (and/or rank). In its turn, this
conclusion makes the metric-compatible connections, maybe, the most effective
mathematical tool for (nonquantum) description of gravitation.

\section*{Acknowledgments}

	The author expresses his gratitude to Nugzar V. Makhaldiani
(Laboratory of Computing Technique and Automation, Joint Institute for
Nuclear Research, Dubna, Russia) for paying his attention to the problem of
signature change in 1994. He thanks Iskra S. Slavova (head of the OBK
Chemical Laboratory, ED~Balkanceramics Ltd., Novi Iskar, Bulgaria) without
whose accidental help in April 1997 this work would not possibly be written.

	This work was partially supported by the National Foundation for
Scientific Research of Bulgaria under Grant No.~F642.

\bibliography{bozhopub,bozhoref}

%
\begin{thebibliography}{10}

\bibitem{sign5}
Hartle J. and Hawking S.
\newblock Wave function of the {U}niverse.
\newblock {\em Phys. Rev. {\textbf{D}}}, 28:2960, 1983.

\bibitem{sign2}
Sakharov A.
\newblock Cosmological transitions with changes in the signature of the metric.
\newblock {\em Sov. Phys. JETP}, 60:214, 1984.

\bibitem{sign7}
Wheeler~J. A.
\newblock On the nature of quantum geometrodynamics.
\newblock {\em Ann. Phys.}, 2:604--614, 1957.

\bibitem{sign8}
Geroch~R. P.
\newblock Topology in general relativity.
\newblock {\em J. Math. Phys.}, 8:782--786, 1967.

\bibitem{sign16}
Gibbons~G. W. and Hartle~J. B.
\newblock Real tunneling geometries and the large-scale topology of the
  {U}niverse.
\newblock {\em Phys. Rev. {\textbf{D}}}, 42(8):2458--2468, 1990.

\bibitem{sign17}
Gary T.
\newblock Topology change in classical and quantum gravity.
\newblock Preprint UCSBTH-90-51, 1990.

\bibitem{sign9}
Gibbons~G. W. and Hawking~S. W.
\newblock Selection rules for topology change.
\newblock {\em Comm. Math. Phys.}, 148:345--352, 1992.

\bibitem{sign14}
Hayward~S. A.
\newblock Signature change in general relativity.
\newblock {\em Class. Quantum Grav.}, 9:1951--1962, 1992.

\bibitem{sign18}
Gibbons~G. W.
\newblock Topology and topology changing in general relativity.
\newblock Preprint DAMTP/R-92/29, 1992.

\bibitem{sign1}
Alty~L. J.
\newblock Kleinian signature change.
\newblock Preprint DAMTP R94/18, May 1984.

\bibitem{sign4}
Dray T., Manoque C., and Tucker R.
\newblock Particle production from signature change.
\newblock {\em General Relativity and Gravitation}, 23:967, 1993.

\bibitem{sign3}
Dray T., Manoque~C. A., and Tucker~R. W.
\newblock The scalar field equation in the presence of signature change.
\newblock {\em Phys. Rev. {\textbf{D}}}, 48:2587--2590, 1993.
\newblock (LANL xxx archive server, E-print No. gr-qc/9303002).

\bibitem{sign23}
Hayward~S. A.
\newblock Junction conditions for signature change.
\newblock LANL xxx archive server, E-print No. gr-qc/9303034, 1993.

\bibitem{sign10}
Gibbons~G. W. and Hawking~S. W.
\newblock Kinks and topology change.
\newblock {\em Phys. Rev. Lett.}, 69:1719--1721, 1992.

\bibitem{sign12}
Alty~L. J.
\newblock The generalized {G}auss-{B}onnet-{C}hern theorem.
\newblock Preprint DAMTP~R94/39, 1994.

\bibitem{sign6}
Alty L.J.
\newblock Binding blocks for topology change.
\newblock Preprint DAMTP R94/38, 1994.

\bibitem{sign30}
Egusquiza~I. L.
\newblock Self-adjoint extensions and signature change.
\newblock {\em Class. Quantum Grav.}, 12:L89--L92, 1995.
\newblock (See also LANL xxx archive server, E-print No. gr-qc/9503015).

\bibitem{sign26}
Embacher F.
\newblock Actions for signature change.
\newblock {\em Phys. Rev. {\textbf{D}}}, 51:6764--6777, 1995.
\newblock (See also LANL xxx archive server, E-print No. gr-qc/9501004 and
  preprint UWThPh-1995-1).

\bibitem{sign13}
Alty~L. J.
\newblock Initial value problems and signature change.
\newblock Preprint DAMTP~R95/1, 1995.

\bibitem{sign21}
Alty~L. J. and Fewster~C. J.
\newblock Initial value problems and signature change.
\newblock {\em Class. Quant. Gravity}, 13:1129, 1996.
\newblock (See also LANL xxx archive server, E-print No. gr-qc/9501026 and
  preprint DAMTP R95/1).

\bibitem{sign20}
Dray T., Ellis G., Helaby Ch., and Manogue C.
\newblock Gravity and signature change.
\newblock {\em General Relativity and Gravitation}, 29:591--597, 1997.
\newblock (LANL xxx archive server, E-print No. gr-qc/9610063).

\bibitem{sign28}
Kriele M.
\newblock Distinguished solutions for discontinuous signature change with weak
  junction conditions.
\newblock LANL xxx archive server, E-print No. gr-qc/9610016, 1996.

\bibitem{sign22}
Hayward~S. A.
\newblock Signature change at material layers and step potentials.
\newblock LANL xxx archive server, E-print No. gr-qc/9509052, 1995.

\bibitem{sign15}
Alty~L. J.
\newblock Kleinian signature change.
\newblock {\em Class. Quantum Grav.}, 11:2523--2536, 1994.

\bibitem{sign19}
Dray T.
\newblock Einstein equations in the presence of signature change.
\newblock {\em J. Math. Phys.}, 37:5627, 1996.
\newblock (LANL xxx archive server, E-print No. gr-qc/9610064).

\bibitem{Greub&et.al.-1}
Greub W., Halperin S., and Vanstone R.
\newblock {\em Connections, Curvature, and Cohomology}, volume~1.
\newblock Academic Press, New York and London, 1972.

\bibitem{Bruhat}
Choquet-Bruhat~Y. at~el.
\newblock {\em Analysis, manifolds and physics}.
\newblock North-Holland Publ.Co., Amsterdam, 1982.

\bibitem{Rosenfel'd}
Rosenfel'd~B. A.
\newblock {\em Non-Euclidean spaces}.
\newblock Nauka, Moscow, 1969.
\newblock (In Russian).

\bibitem{Mathenedia-4}
\newblock {\em Mathematical encyclopedia}, volume~4.
{Vinogradov I. M., chief editor}.
\newblock Soviet encyclopedia, Moscow, 1984.
\newblock  (In Russian).

\bibitem{Tricomi}
Tricomi~F. O.
\newblock {\em On linear partial differential equations}.
\newblock Gostekhizdat, Moscow, 1947.
\newblock (Russian translation from Italian).

\bibitem{MoiseevEI}
Mo{\v\i}seev~E. I.
\newblock {\em Equations of mixed type with spectral parameter}.
\newblock Moscow University, Moscow, 1988.
\newblock (In Russian).

\bibitem{FriedrichsKO1958}
Friedrichs~K. O.
\newblock Symmetric positive linear differential equations.
\newblock {\em Comm. Pure Appl. Math.}, 11(3):333--418, 1958.

\bibitem{FriedrichsKO1970}
Friedrichs~K. O.
\newblock {\em Pseudo-differential operators. An introduction}.
\newblock Lecture notes. Courant institute of mathematical sciences, New York,
  1970.

\bibitem{SmirnovMM}
Smirnov~M. M.
\newblock {\em Degenerate elliptic and hyperbolic equations}.
\newblock Nauka, Moscow, 1966.
\newblock (In Russian).

\bibitem{Oleinik&Rodkevich}
Ole{\v\i}nik~O. A. and Rodkevich~E. V.
\newblock {\em Second order equations with non-negative characteristic form}.
\newblock Review of science, Section Mathematics: Mathematical analysis 1969.
  VINITI, Moscow, 1971.
\newblock (In Russian).

\bibitem{BerezanskiiYuM}
Berezanski{\v\i}~Yu. M.
\newblock {\em Expansion of selfadjoint operators over eigenfunctions}.
\newblock Naukova dumka, Kiev, 1965.
\newblock (In Russian).

\bibitem{MikhlinSG}
Mikhlin~S. G.
\newblock {\em Linear partial differential equations}.
\newblock V$\overline{\mathrm{y}}$shaya shkola, Moscow, 1977.
\newblock (In Russian).

\bibitem{Kuzmin'AG}
Kuzmin'~A. G.
\newblock {\em Non-classical equations of mixed type and their application to
  gas dynamics}.
\newblock Leningrad University, Leningrad, 1990.
\newblock (In Russian).

\bibitem{GuChaohao}
Gu~Chaohao and Hong Jiaxing.
\newblock {\em Some developments of the theory of partial differential
  equations of mixed type}, volume 90 of {\em Teubner Texte zur
  Mathematik}, pages 120--135.
\newblock Teubner Verlag Geselschaft, Leipzig, 1986.

\bibitem{f-LTP-general}
Iliev~B. Z.
\newblock Linear transports along paths in vector bundles. {I}.~{G}eneral
  theory.
\newblock JINR Communication E5-93-239, Dubna, 1993.

\bibitem{f-LTP-metrics}
Iliev~B. Z.
\newblock Linear transports along paths in vector bundles. {IV}.~{C}onsistency
  with bundle metrics.
\newblock JINR Communication E5-94-17, Dubna, 1994.

\bibitem{f-TP-parallelT}
Iliev~B. Z.
\newblock Transports along paths in fibre bundles. {II}.~{T}ies with the theory
  of connections and parallel transports.
\newblock JINR Communication E5-94-16, Dubna, 1994.

\bibitem{K&N-1}
Kobayashi S. and Nomizu K.
\newblock {\em Foundations of Differential Geometry}, volume~I.
\newblock Interscience Publishers, New York-London, 1963.

\bibitem{K&N-2}
Kobayashi S. and Nomizu K.
\newblock {\em Foundations of Differential Geometry}, volume~II.
\newblock Interscience Publishers, New York-London-Sydney, 1969.

\bibitem{Lankaster-matrices}
Lankaster P.
\newblock {\em Theory of matrices}.
\newblock Academic Press, New York-London, 1969.

\bibitem{Gantmacher-matrices-I}
Gantmacher~F. R.
\newblock {\em The theory of matrices}, volume one.
\newblock Chelsea Pub. Co., New York, N.Y., 1960.
\newblock (Translation from Russian. The right English transliteration of the
  author's name is Gantmakher).

\bibitem{Waerden-algebra-II}
van~der Waerden B.~L.
\newblock {\em Algebra II}.
\newblock Springer-Verlag, Berlin-Heidelberg-New York, fifth edition, 1967.
\newblock (In German).

\bibitem{Gantmacher-matrices-II}
Gantmacher~F. R.
\newblock {\em The theory of matrices}, volume two.
\newblock Chelsea Pub. Co., New York, N.Y., 1960 (reprinted 1964).
\newblock (Translation from Russian. The right English transliteration of the
  author's name is Gantmakher).

\bibitem{Gronwald/Hehl-96}
Gronwald F. and Hehl~F. W.
\newblock On gauge aspects of gravity.
\newblock In Bergmann~P. G., de~Sabbata~V., and Treder H.-J., editors, {\em
  Quantum Cosmology}, Proc. of the 14-th Course of the School of Cosmology and
  Gravitation, pages 148--198, Erice, Italy, May 1995, 1996. World Scientific,
  Singapore.
\newblock (LANL xxx archive server, E-print No. gr-qc/9602013).

\bibitem{MTW}
Misner~C. W., Thorne~K. S., and Wheeler~J. A.
\newblock {\em Gravitation}.
\newblock W. H. Freeman and Company, San Francisco, 1973.

\bibitem{Bishop/Crittenden}
Bishop~R. L. and Crittenden~R. J.
\newblock {\em Geometry of Manifolds}.
\newblock Academic Press, New York-London, 1964.

\bibitem{f-PE-P?}
Iliev~B. Z.
\newblock Is the principle of equivalence a principle?
\newblock {\em Journal of Geometry and Physics}, 24(3):209--222, 1997.

\bibitem{Hehl-95-review}
Hehl~F. W., McCrea~J. D., Mielke~E. W., and Ne'eman Y.
\newblock Metric-affine gauge theory of gravity: field equations, {N}oether
  identities, world spinors, and breaking of dilation invariance.
\newblock {\em Phys. Rep.}, 258(1 \& 2):1--171, July 1995.

\bibitem{sign24}
Embacher F.
\newblock Dimensionality, topology, energy, the cosmological constant, and
  signature change.
\newblock {\em Class. Quantum Grav.}, 13:921, 1996.
\newblock (See also LANL xxx archive server, E-print No. gr-qc/9504040 and
  preprint UWThPh-1995.11).

\bibitem{sign25}
Embacher F.
\newblock Space-time dimension, Euclidean action and signature change.
\newblock Preprint UWThPh-1995.28, 1995.
\newblock (See also LANL xxx archive server, E-print No. gr-qc/9507041).

\bibitem{sign27}
Embacher F.
\newblock Signature change induces compactification.
\newblock {\em Phys. Rev. {\textbf{D}}}, 52:2150, 1995.
\newblock (See also LANL xxx archive server, E-print No. gr-qc/9410012 and
  preprint UWThPh-1994-47).

\end{thebibliography}
%
\bibliographystyle{unsrt}

\end{document}